\documentclass[
preprintnumbers,nofootinbib,aps,prd,floatfix]{revtex4-2}
\pdfoutput=1 

\usepackage{subfigure,graphicx,amsmath,amssymb,hyperref}
\usepackage{youngtab}

\usepackage{float}

\newcommand{\bea}{\begin{eqnarray}}
\newcommand{\eea}{\end{eqnarray}}

\def\beq#1\eeq{\begin{align}#1\end{align}}
\def\beqnn#1\eeq{\begin{align*}#1\end{align*}}

\usepackage[usenames,dvipsnames]{xcolor}
\definecolor{darkgreen}{rgb}{0,0.5,0}

\begin{document}
\preprint{UCI-HEP-TR 2022-25}
\title{Invariant Tensors for $SU(3)$ with Adjoints}

\author{Arvind Rajaraman}
\email{arajaram@uci.edu}
\affiliation{Department of Physics and Astronomy,\\
University of California, Irvine, CA 92697-4575 USA
}

\begin{abstract}
We develop a method for finding the independent 
invariant tensors of a gauge theory.
Our method uses a theorem relating  invariant tensors and 
constant configurations in field space.
We apply our method to an $SU(3)$ gauge theory with matter in the adjoint representation, and find the independent 
invariant tensors of this theory.
\end{abstract}

\maketitle


\section{Introduction}
A gauge theory is specified by the gauge group and 
the representation of the matter fields
under the gauge group,  
 but all observables, including the physical spectrum, 
 are gauge invariant combinations of fields. The structure of
 these objects is found by contracting the
 gauge-covariant fields with 
 {\it invariant tensors} to form invariant objects.
 
 For example, the group $SO(3)$ has one invariant tensor with two indices ($\delta^{ij}$), and an  invariant tensor  with three indices  ($\epsilon^{ijk}$).  One can contract 
 two fields in the fundamental representation with $\delta^{ij}$, or  three fields in the fundamental representation with $\epsilon^{ijk}$, and these give invariants of $SO(3)$. Indeed, this is a complete set of basis invariants for a $SO(3)$ theory with fields in the fundamental representation.
Powerful calculational methods have been developed for these invariant tensors in this and other theories (e.g.  the method of birdtracks~\cite{Cvitanovic:2008zz}). 
Using such methods, it is possible to calculate  many relations between these tensors, such as contractions of tensors.

However, one question that is not answered by these methods is the question: how many independent tensors exist in a given theory? That is, given a symmetry group and a set of fields in various representations, how many independent contractions can we have between the invariant tensors and the fields?

A brief example will suffice to show the kinds of difficulties that may occur.
As mentioned above, it is known that in the group $SO(3)$, the invariant tensors are $\delta^{ij}, \epsilon^{ijk}$. These
tell us all the invariants in a theory where all fields are in the fundamental representation.
But in a theory with $SO(3)$ gauge symmetry and
with fields 
 $V^{(1)}_{ij}, V^{(2)}_{ij}..$ in the symmetric tensor representation, one can already produce an infinite set of 
 invariants by contracting
an arbitrarily long sequence $V^{(1)}_{ij}V^{(2)}_{jk}..V^{(N)}_{li}$ (and there may exist
further invariants involving epsilon tensors).  

However, general theorems, due to Hilbert, assure us that all the members of this infinite set of invariants are generated by a finite set of basis invariants. (The set of invariants is a ring, and this basis is sometimes referred to as the Hilbert basis. {For some of the mathematical 
literature relevant to invariant theory see:~\cite{Weyl:1946}--
\cite{Gufan:2001}.}) From a physical point of view, this is also expected since it would be surprising if a finite number of degrees of freedom encoded in the symmetric tensors could yield an infinite number of composites.
Furthermore, for certain physical applications, it is important to know the Hilbert basis. For example, duality relations between supersymmetric theories~\cite{Seiberg:1995pq} require a knowledge of the chiral ring of supersymmetric invariants. More generally, any confining theory will have a spectrum of composites which will correspond to these independent invariants.

To look for this basis, one might try to find relations  between products of invariant tensors which allow them to be reduced. In practice, these are very difficult to find~\cite{Pouliot:2001iw}, and we will not take this approach. Instead we will
apply a theorem that directly allows to find a set of basis invariants. This theorem states that there is  a one-to-one correspondence between invariant operators and gauge-fixed configurations of the fields (a precise statement is in \cite{Luty:1996sd}). Finding all  gauge fixed configurations would then tell us the full set of invariants.

The logic of the theorem is simple. 
An arbitrary constant configuration of the fields can be gauge-fixed; that is, one can use gauge transformations to eliminate some parameters of the configuration until the gauge transformations are exhausted (this assumes that both the fields and the gauge symmetries are complexified). The remaining parameters are all, by construction, gauge-invariant operators. In this gauge, there is therefore an obvious one-to-one correspondence between parameters of the configuration and the gauge-invariant operators. This will then continue to hold in any gauge.

The approach to finding the basis set of invariants would then seem straightforward. We should consider all possible configurations of the fields and gauge-fix each configuration so that the gauge symmetry is broken. In this specific gauge, each remaining parameter of the configuration would be  an invariant tensor. Finally we would then find a set of invariant tensors  that would reduce to these tensors in the special gauge. 

While  straightforward, there are certain challenges with this approach~\cite{Berger:2018dxg}, which we will illustrate further below. In the first place, it is not trivial to completely gauge fix the group symmetry. Different configurations have different gauge fixing procedures, and it is not easy to categorize these. We will find that it is easier to partially gauge fix the symmetry, leaving an unbroken part of the symmetry, and to find the Hilbert basis of the reduced symmetry, rather than a completely broken symmetry.
Secondly, there is no obvious way to remove the gauge-fixing. We therefore need to explore all invariants of the full theory, and compare them case by case to the gauge-fixed invariants.

We then show the practicality of this approach
by studying a theory with $SU(3)$ gauge group and fields in the adjoint representation. Partial results for the Hilbert basis have been obtained before for these theories~\cite{Dittner:1971fy}, \cite{Dittner:1972hm}.
These theories are simple enough that our methods can be applied fairly straightforwardly, and they also have relation to phenomenological theories. Here we are able to explicitly find the basis set of  invariants for this  theory.

We establish our notation in the following section, and begin the gauge fixing procedure. Generically,
this gauge fixing leaves a residual $U(1)\times U(1)$ symmetry, for which we are able to find the gauge-fixed invariants; i.e. the combinations of fields invariant under the residual symmetry. We further identify a permutation  symmetry in the following section, which we use to further characterize the gauge-fixed invariants. 
There is furthermore a special set of choices for which the gauge-fixing 
leaves a residual $SU(2)\times U(1)$ symmetry. The invariants for this theory are identified in section IV. 

Finally, we look for the  invariants of the full $SU(3)$ symmetry in section V. The program LIE~\cite{LIE} proved invaluable in this effort. The invariants of the theory are identified and we further check in an appendix that all gauge fixed invariants can be reproduced from the list of $SU(3)$ invariants, thereby establishing that we have found the full set of independent invariants.

\section{SU(3)  adjoints: The gauge-fixed field space }

\subsection{Notation}
We consider a theory with a gauge symmetry of $SU(3)$, and  $N$ matter fields transforming in the adjoint of $SU(3)$.
We denote the adjoints as $X_{i\dot{j}}^I$
where $i,\dot{j} $ (the gauge indices) are
the fundamental and antifundamental indices of $SU(3)$, and $I=1..N$  labels the
different fields. We shall call the $I$ index a flavor index in analogy with the flavors in particle physics.
We shall   use lower
case indices for gauge indices and upper case for the flavor indices. 

 The general constant configuration of these adjoints  is of the form
 \bea
X^{I}=
\left(\begin{array}{ccc}X^{I}_{1\dot{1}}&X^{I}_{1\dot{2}}&
X^{I}_{1\dot{3}}\\
X^{I}_{2\dot{1}}&X^{I}_{2\dot{2}}&X^{I}_{2\dot{3}}\\
X^I_{3\dot{1}}&X^{I}_{3\dot{2}}&X^{I}_{3\dot{3}}\end{array}
\right)
\label{XIform}
\eea

Under a gauge transformation,
\bea
\delta X^{I}=[i\alpha_at^a,X^{I}]
\eea
where $t^a$ are the Gell-Mann matrices.

We now try to gauge fix the symmetry. Using the 
$SU(3)$ gauge transformations, we can bring one  adjoint (which we label $X^0$) to the form
\bea
X^{0}=
\left(\begin{array}{ccc}a&0&
0\\
0&b&0\\
0&0&c
\end{array}
\right)
\label{X0form}
\eea
with $a+b+c=0$. 

This gauge fixing  partially breaks the
$SU(3)$ symmetry.
We have a preserved
continuous symmetry of $U(1)\times U(1)$ generated by
\bea
t^{3}=
\left(\begin{array}{ccc}1&0&
0\\
0&-1&0\\
0&0&0
\end{array}
\right)\qquad t^{8}=
\left(\begin{array}{ccc}1&0&
0\\
0&1&0\\
0&0&-2
\end{array}
\right)
\eea
 $X^0$ (and hence $a,b,c$) are invariant under the action of $U(1)\times U(1)$.

We could then try to gauge fix these residual symmetries (using the other adjoints) to find the gauge invariant parameters. Unfortunately, this cannot be done in an universal way. For example, if we are on a point in configuration space where only $X_{12}$ is nonzero, we could attempt to break the symmetry further by gauge-fixing   one of the fields. On the other hand, if we  are on a point in configuration space where no  $X_{12}$ is nonzero, we would need to consider other fields. In general,  the gauge fixing can be done in many different ways, and it becomes difficult to keep track of the options. This will of course become exponentially worse for more complicated symmetries or larger representations.

\subsection{Gauge-fixed $U(1)\times U(1)$ invariants}

We therefore take a different approach.
Since we have a one-to-one correspondence between the configuration parameters and the gauge-invariant operators, we can just as well  directly look for the $U(1)\times U(1)
$ gauge invariant operators describing the configuration
(\ref{XIform},\ref{X0form}). Since these are found after gauge fixing one adjoint, these operators will in general not be manifestly $SU(3)$ invariant. We will then  
 need to find fully $SU(3)$ invariant operators that are in one-to-one correspondence with these   
 operators with $U(1)\times U(1)$  gauge invariance.

Under the    $U(1)\times U(1)$ transformations, we have 
\bea
\delta_{t^3} X^{I}=
\delta_{t^3}\left(\begin{array}{ccc}X^{I}_{1\dot{1}}&X^{I}_{1\dot{2}}&
X^{I}_{1\dot{3}}\\
X^{I}_{2\dot{1}}&X^{I}_{2\dot{2}}&X^{1}_{2\dot{3}}\\
X^I_{3\dot{1}}&X^{I}_{3\dot{2}}&X^{I}_{3\dot{3}}\end{array}
\right)=
i\alpha_3\left(\begin{array}{ccc}0&2X^{I}_{1\dot{2}}&
X^{I}_{1\dot{3}}\\
-2X^{I}_{2\dot{1}}&0&-X^{1}_{2\dot{3}}\\
-X^I_{3\dot{1}}&X^{I}_{3\dot{2}}&0\end{array}
\right)
\\
\delta_{t^8} X^{I}=
\delta_{t^8}\left(\begin{array}{ccc}X^{I}_{1\dot{1}}&X^{I}_{1\dot{2}}&
X^{I}_{1\dot{3}}\\
X^{I}_{2\dot{1}}&X^{I}_{2\dot{2}}&X^{1}_{2\dot{3}}\\
X^I_{3\dot{1}}&X^{I}_{3\dot{2}}&X^{I}_{3\dot{3}}\end{array}
\right)
=i\alpha_8\left(\begin{array}{ccc}0&0&
3X^{I}_{1\dot{3}}\\
0&0&3X^{1}_{2\dot{3}}\\
-3X^I_{3\dot{1}}&-3X^{I}_{3\dot{2}}&0\end{array}
\right)
\eea
so that we can identify the  charges
of each component of the adjoint under the $U(1)\times U(1)$ as
\begin{table}[ht]
\centering 
\begin{tabular}{|c| c| c|c|c|c|c|c|c|c| } 
\hline 
Field &
$X^{I}_{1\dot{1}}$ &
$X^{I}_{1\dot{2}}$ &
$X^{I}_{1\dot{3}}$ &
$X^{I}_{2\dot{1}}$ &
$X^{I}_{2\dot{2}}$ &
$X^{I}_{2\dot{3}}$ &
$X^{I}_{3\dot{1}}$ &
$X^{I}_{3\dot{2}}$ &
$X^{I}_{3\dot{3}}$ 
\\
\hline 
$U(1)_1$ 
&0&2&1&-2&0&-1&-1&1&0
\\
\hline 
 $U(1)_2$  
 &0&0&3&0&0&3&-3&-3&0\\ [0.5ex] 
\hline 
\end{tabular}
\caption{Charges of fields under $U(1)\times U(1)$}
\end{table}

It is straightforward to find all the combinations of fields invariant under the $U(1)\times U(1)$. These are generated by a small set of operators. This set of operators, along with the possible symmetries that can be associated with the flavor indices, is shown in table \ref{U1invts}. 
We shall refer to these as the {\it gauge-fixed invariants} at the generic parameter point.
\begin{table}[h]

\centering
\begin{tabular}{|c| c| c| c| c|} 
\hline\hline 
Operator & Symmetries of Flavor Indices   \\ [0.5ex] 
\hline 
$X^{I}_{1\dot{1}}
,
X^{I}_{2\dot{2}}
,
X^{I}_{3\dot{3}}$& \rule{0pt}{1.2em}$\begin{array}{c}\yng(1)\end{array}$ 
\\
\hline
$X^{I}_{2\dot{3}}X^{J}_{3\dot{2}}
,
X^{I}_{3\dot{1}}X^{J}_{1\dot{3}},
X^{I}_{1\dot{2}}X^{J}_{2\dot{1}}$ & \rule{0pt}{1.8em}$\begin{array}{c}\yng(2)\end{array}$\ 
+ $\begin{array}{c}\yng(1,1)\end{array}$
\\
\hline
$X^{I}_{1\dot{2}}X^{J}_{2\dot{3}}X^{K}_{3\dot{1}}
,X^{I}_{2\dot{1}}X^{J}_{3\dot{2}}X^{K}_{1\dot{3}}$ & \rule{0pt}{2.2em}$\begin{array}{c}\yng(3)\end{array}$\ 
+$\begin{array}{c}\yng(2,1)\end{array}$+$\begin{array}{c}\yng(1,1,1)\end{array}$ 
\\
\hline
\end{tabular}
\caption{This table shows the combinations of fields invariant under the $U(1)\times U(1)$ subgroup. The second column shows the various flavor symmetries possible for each combination.}
\label{U1invts}
\end{table}

\section{The permutation symmetry}

While our goal is to find the $SU(3)$ invariant operators that match on to the operators found above, there is an intermediate step that is extremely useful and informative, where we find operators that are invariant both under the $U(1)\times U(1)$ symmetry found above, as well as a permutation symmetry. In this section, we describe this intermediate step.

\subsection{ The action of the permutations on the gauge indices}

The  discrete symmetry we will consider  can be thought of as a permutation of the gauge indices (i.e. of (1,2,3) and $\dot{1}, \dot{2}, \dot{3}$). This then translates to an action on the 
fields 
which (among other actions) permutes the three diagonal entries $(a,b,c)$. This permutation is not in general part of the full $SU(3)$ gauge symmetry, because a $SU(3)$ invariant with a factor of an epsilon tensor  is odd under this permutation. However, since the $SU(3)$ adjoint has an even number of indices, any invariant will have an even number of epsilon tensors, and will therefore be invariant under the permutation. For this theory (i.e. a $SU(3)$ gauge theory with all fields in the adjoint represenation), therefore, the permutation is a symmetry.

Since this symmetry acts on the three eigenvalues $(a,b,c)$ by permuting them, the symmetry is (spontaneously) broken by the gauge fixing once  a particular choice of ordering of the eigenvalues is chosen.
The operators found above on the gauge-fixed field space therefore also do not exhibit this symmetry. 
Since we know that the $SU(3)$ invariants must be invariant under the permutations, it is necessary to combine the $U(1)\times U(1)$ invariant operators found above with the eigenvalues in such a way that the combination is permutation invariant.

Making the discrete symmetry manifest can  therefore be used as an important step towards making the full $SU(3)$  manifest.

\subsection{ The relevant representations of the  permutation group}

We shall treat the permutations as being 
 generated by two elements. The first is a cyclic permutation $C$ which acts on the gauge indices as $C(1)=2, C(2)=3, C(3)=1$.  
The second is an exchange permutation $E$
 that will be taken to act on the gauge indices as $ E(1)=1, E(2)=3, E(3)=2$.

More generally, we define the 3 representation as a triplet of objects
$V=(\alpha,\beta,\gamma)$
with
\bea
C(\alpha)=\beta, C(\beta)=\gamma, C(\gamma)=\alpha
\nonumber
\\
E(\alpha)=\alpha, E(\beta)=\gamma, E(\gamma)=\beta
\eea

The other representation which will occur in the following analysis is a triplet of objects
$V'=(\alpha',\beta',\gamma')$
with
\bea
C(\alpha')=\beta', C(\beta')=\gamma', C(\gamma')=\alpha'
\nonumber
\\
E(\alpha')=-\alpha', E(\beta')=-\gamma', E(\gamma')=-\beta'
\eea
We shall denote this in the following as  a 3' representation. 

We should note that both these representations are reducible; we shall not need to worry about this here.

\subsection{ The action of the permutations on the invariants}

We now  classify the invariants under the action of the permutations.

The three eigenvalues of $X^0$ are acted on as
\bea
C(a)=b, C(b)=c, C(c)=a\qquad 
E(a)=a, E(b)=c, E(c)=b
\eea
and therefore are in a 3 representation. We denote this as
the triplet $V^0=(a,b,c)$.

Similarly, we find
 \bea
V^{I}&&=(X^{I}_{1\dot{1}}
,
X^{I}_{2\dot{2}}
,
X^{I}_{3\dot{3}})
\nonumber
\\
V^{IJ}&&=
(X^{\{I}_{2\dot{3}}X^{J\}}_{3\dot{2}}
,
X^{\{I}_{3\dot{1}}X^{J\}}_{1\dot{3}}
,
X^{\{I}_{1\dot{2}}X^{J\}}_{2\dot{1}})
\eea
both transform in the 3 representation, while
\bea
V^{'IJ}=
X^{[I}_{2\dot{3}}X^{J]}_{3\dot{2}}
,
X^{[I}_{3\dot{1}}X^{J]}_{1\dot{3}}
,
X^{[I}_{1\dot{2}}X^{J]}_{2\dot{1}}
\eea
transforms in the $3'$ representation.

The operators with three fields are organized as
\bea
i_1^{IJK}=X^{I}_{1\dot{2}}X^{J}_{2\dot{3}}X^{K}_{3\dot{1}}
+
X^{I}_{1\dot{3}}X^{J}_{3\dot{2}}X^{K}_{2\dot{1}}
\nonumber
\\
i_2^{IJK}=
X^{I}_{2\dot{3}}X^{J}_{3\dot{1}}X^{K}_{1\dot{2}}
+
X^{I}_{2\dot{1}}X^{J}_{1\dot{3}}X^{K}_{3\dot{2}}
\nonumber
\\
i_3^{IJK}=
X^{I}_{3\dot{1}}X^{J}_{1\dot{2}}X^{K}_{2\dot{3}}
+
X^{I}_{3\dot{2}}X^{J}_{2\dot{1}}X^{K}_{1\dot{3}}
\nonumber
\\
i_4^{IJK}=X^{I}_{1\dot{2}}X^{J}_{2\dot{3}}X^{K}_{3\dot{1}}
-
X^{I}_{1\dot{3}}X^{J}_{3\dot{2}}X^{K}_{2\dot{1}}
\\
i_5^{IJK}=
X^{I}_{2\dot{3}}X^{J}_{3\dot{1}}X^{K}_{1\dot{2}}
-
X^{I}_{2\dot{1}}X^{J}_{1\dot{3}}X^{K}_{3\dot{2}}
\nonumber
\\
i_6^{IJK}=
X^{I}_{3\dot{1}}X^{J}_{1\dot{2}}X^{K}_{2\dot{3}}
-
X^{I}_{3\dot{2}}X^{J}_{2\dot{1}}X^{K}_{1\dot{3}}
\nonumber
\eea

We then find that 

\bea
V^{IJK}=(i_1^{IJK}+i_1^{IKJ},i_2^{IJK}+i_2^{IKJ},i_3^{IJK}+i_3^{IKJ})
\nonumber
\\
W^{IJK}=(i_1^{IJK}-i_1^{IKJ},i_2^{IJK}-i_2^{IKJ},i_3^{IJK}-i_3^{IKJ})
\eea
transform in a 3 representation, while
\bea
V^{'IJK}=(i_4^{IJK}+i_4^{IKJ},i_5^{IJK}+i_5^{IKJ},i_6^{IJK}+i_6^{IKJ})
\nonumber
\\
W^{'IJK}=(i_4^{IJK}-i_4^{IKJ},i_5^{IJK}-i_5^{IKJ},i_6^{IJK}-i_6^{IKJ})
\eea
transform in a $3'$ representation.

\subsection{ Invariants of the permutation groups}


The full invariants of the theory must be invariant under the $S_3$ permutation group. The reason that this is not evident in the $U(1)\times U(1)$ invariants is that the symmetry has been broken by the choice of a specific $X^0$. The $U(1)\times U(1)$ invariants above must therefore be combined with the vector representation  $V^0=(a,b,c)$  to form invariants under $S_3$. 
Since we want our invariants to be in one-to-one correspondence with the $U(1)\times U(1)$ invariants in the table II,  we should consider combinations which are linear in the gauge-fixed invariants but can contain arbitrary (as it turns out, up to cubic) powers of $V^0$.

To find the invariants under $S_3$, we first find the invariants under the cyclic permutation, and then impose the exchange permutation.

For a vector representation,
we 
form the combinations \bea
v=\alpha+\omega \beta+\omega^2 \gamma
\qquad 
w=\alpha+\omega^2 \beta+\omega \gamma
\eea
which have charge $\omega, \omega^2$ under the cyclic permutation, and are interchanged by the exchange action.
Similarly we  define
\bea
v_0=a+\omega b+\omega^2 c
\qquad
w_0=a+\omega^2 b+\omega c
\eea

Singlets under C are found by taking a product of a charge $\omega$ and a charge  $\omega^2$, or three charges $\omega$, or three charges $\omega^2$. Further imposing invariance under the  
exchange action yields the $S_3$ invariants
\bea
Tr(V)&\equiv& \alpha+\beta+\gamma
\nonumber
\\
V^0\cdot V&\equiv& a\alpha+b\beta+c\gamma
\\
(V^0)^2\cdot  V&\equiv&   a^2\alpha+b^2\beta+c^2\gamma
\nonumber
\eea
Similarly, for the $3'$, we
find the $S_3$ invariants
\bea
(V^0)^3\diamond  Tr(V')&\equiv& ( a^2b- a^2c -ab^2+a c^2+b^2 c- b c^2)(\alpha'+\beta'+\gamma')
\nonumber
\\
V_0\diamond V'&\equiv& (
b\alpha'+c\beta'+a\gamma')-(c\alpha'+a\beta'+b\gamma'
)
\\
(V^0)^2\diamond V'&\equiv& (
b^2\alpha'+c^2\beta'+a^2\gamma')-(c^2\alpha'+a^2\beta'+b^2\gamma')
\nonumber
\eea

\subsection{ The invariants under $U(1)\times U(1)\times S_3$ }

Putting everything together, the invariants under the product of the continuous symmetry $U(1)\times U(1)$
and the discrete symmetry $S_3$ are listed in the table below.

\begin{table}[H]
\centering 

\begin{tabular}{|c| c| c|c|c|c| } 

\hline 
Invariant& Flavor Symmetry 
&
Invariant& Flavor Symmetry 
\\
\hline 
$V_0\cdot V^{I}$
&\rule{0pt}{1.2em}   
$ \begin{array}{c}\yng(1)\end{array}$
&
$Tr(V^{IJK})$
& \rule{0pt}{1.2em}  $\begin{array}{c} \yng(3)\end{array}$
\\
\hline 
$  V_0^2\cdot V^{I}$
&\rule{0pt}{1.2em}   $\begin{array}{c} \yng(1)\end{array}$
&
$ Tr(W^{IJK})$
&\rule{0pt}{2.2em}   $\begin{array}{c} \yng(1,1,1)\end{array}$
\\
\hline 
$Tr(V^{IJ})$
& \rule{0pt}{1.2em}  
$ \begin{array}{c}\yng(2)\end{array}$
&
$V_0\cdot V^{IJK}, V_0\cdot W^{IJK}$
&\rule{0pt}{1.8em}   $\begin{array}{c} \yng(2,1)\end{array}$
\\
\hline 
$V_0\cdot V^{IJ}$
&\rule{0pt}{1.2em}   $\begin{array}{c} \yng(2)\end{array}$
&
$  V_0^2\cdot V^{IJK}, V_0^2\cdot W^{IJK}$ 
&\rule{0pt}{1.8em}   $\begin{array}{c} \yng(2,1)\end{array}$
\\
\hline 
$  V_0^2\cdot V^{IJ}$
&\rule{0pt}{1.2em}   $\begin{array}{c} \yng(2)\end{array}$
&
$V_0\diamond  V^{'IJK}, V_0\diamond  W^{'IJK}$
&\rule{0pt}{1.8em}   $\begin{array}{c} \yng(2,1)\end{array}$

\\
\hline 
$V_0\diamond  V^{'IJ}$
&\rule{0pt}{1.8em}   $\begin{array}{c} \yng(1,1)\end{array}$
&
$  V_0^2\diamond  V^{'IJK},  V_0^2\diamond  W^{'IJK}$
&\rule{0pt}{1.8em}   $\begin{array}{c} \yng(2,1)\end{array}$

\\
\hline 
$  V_0^2\diamond  V^{'IJ}$
& \rule{0pt}{1.8em}  $\begin{array}{c} \yng(1,1)\end{array}$

&

$  V_0^3\diamond  Tr (V^{'IJK})$
&\rule{0pt}{1.2em}   $\begin{array}{c} \yng(3)\end{array}$

\\
\hline 
$  V_0^3\diamond  Tr (V^{'IJ})$
&\rule{0pt}{1.8em}   $\begin{array}{c} \yng(1,1)\end{array}$
&
$  V_0^3\diamond  Tr (W^{'IJK})$
& \rule{0pt}{2.2em}  $\begin{array}{c} \yng(1,1,1)\end{array}$

\\
\hline 
\end{tabular}
\caption{This table shows the combinations of fields invariant under the  $U(1)\times U(1)\times S_3$ subgroup. The second column shows the various flavor symmetries possible for each combination. }
\label{genericgaugefixedinvts}
\end{table}

\section{The Point of Enhanced symmetry}

While the analysis above yields the invariants at generic points in parameter space, there are special points that needs to be considered separately. These are  where the symmetry is enhanced; this can happen when two eigenvalues of $X^0$ coincide  e,g. when   $b=c$.

At this point, 
the symmetry is enhanced;
the unbroken symmetry generators are 
\bea
t^{1}=
\left(\begin{array}{ccc} 
0&0&0\\
0&0&1\\
0&1&0
\end{array}
\right)\qquad 
t^{2}=
\left(\begin{array}{ccc} 
0&0&0\\
0&0&-i\\
0&i&0
\end{array}
\right)\qquad 
t^{3}=
\left(\begin{array}{ccc} 
0&0&0\\
0&1&0\\
0&0&-1
\end{array}
\right)\qquad 
t^{8}=
\left(\begin{array}{ccc} 
-2&0&0\\
0&1&0\\
0&0&1
\end{array}
\right)
\eea
and the continuous symmetry is enhanced to $SU(2) \times U(1)$. 
The invariants therefore need to  be invariant under  
$SU(2) \times U(1)$.

In this case, the theory is simple enough that this can be done directly.
The $SU(3)$ adjoint  decomposes
as an adjoint of $SU(2)$,
\bea
a^{I}=\left(\begin{array}{ccc}{X^{I}_{2\dot{2}}-X^{I}_{3\dot{3}}\over 2}&X^{I}_{2\dot{3}}\\
X^{I}_{3\dot{2}}&{X^{I}_{3\dot{3}}-X^{I}_{2\dot{2}}\over 2}\end{array}
\right)
\eea
two doublets,
\bea
q^I_a=(X^I_{1\dot{2}}, X_{1\dot{3}})
\qquad 
Q^J_{\dot a}=(X^J_{2\dot{1}}, X_{3\dot{1}})
\eea
and 
a scalar
\bea
s^{I}={X^{I}_{2\dot{2}}+X^{I}_{3\dot{3}}\over 2}
\eea


Under the unbroken $U(1)$, the $q, Q$ have charges $3, -3$ respectively, while the scalar and adjoint are neutral.

\begin{table}[t]
\centering 
\begin{tabular}{|c| c| c| c| c|} 
\hline\hline 
Invariant & Flavor Symmetry    \\ [0.5ex] 
\hline 
$s^{I}$& \rule{0pt}{1.2em}$\begin{array}{c}\yng(1)\end{array}$ 

\\
\hline
$q^{I}.Q^{J}$ & \rule{0pt}{1.8em}$\begin{array}{c}\yng(2)\end{array}$\ 
+$\begin{array}{c}\yng(1,1)\end{array}$

\\
\hline
$a^K_i a^L_i$
 &\rule{0pt}{1.2em} $\begin{array}{c}\yng(2)\end{array}$

\\
\hline
$q^{I}\sigma^i Q^{J}a^K_i$ & \rule{0pt}{2.3em}$\begin{array}{c}\yng(3)\end{array}$\ 
+$\begin{array}{c}\yng(2,1)\end{array}$+$\begin{array}{c}\yng(1,1,1)\end{array}$

\\
\hline
$\epsilon^{ijk}a^K_i a^L_ja^M_k$ & \rule{0pt}{2.3em}
$\begin{array}{c}\yng(1,1,1)\end{array}$

\\
\hline
$\epsilon^{ijk}q^{I}\sigma^{i}  Q^{J} a^K_j a^L_k $ & \rule{0pt}{3.0em}
$\begin{array}{c}\yng(3,1)\end{array}+\begin{array}{c}\yng(2,2)\end{array}+\begin{array}{c}\yng(2,1,1)\end{array}+\begin{array}{c}\yng(1,1,1,1)\end{array}$
\\
\hline
\end{tabular}
\caption{This table shows the combinations of fields which are invariant under the  enhanced symmetry subgroup $SU(2)\times U(1)$. The second column shows the possible symmetries of the flavor indices.}
\label{enhancedgaugefixedinvts}
\end{table}

The $U(1)$ invariants are the scalar, the adjoint $a^i$, and  $q^I. Q^J\equiv q^I_a
Q^J_{\dot a}$, and, $q^I.\sigma^i. Q^J\equiv q^I_a\sigma^i_{\dot{a}b}
Q^J_{\dot b}$.
These are scalars and fundamentals under $SO(3)$. The invariant tensors of $SO(3)$ with fundamentals are known to be  $\delta_{ij},\epsilon_{ijk}$, which allows us to form the complete set of invariants of this theory by contracting the invariant tensors with the fundamentals.

Using the Fierz identities
\bea
\sigma^i_{\dot{a}b}\sigma^i_{\dot{c}d}
=-\delta_{\dot{a}b}\delta_{\dot{c}d}
+2\delta_{\dot{a}d}\delta_{\dot{c}b}
\eea
helps us to reduce some of these invariants to products of smaller invariants;
the remaining  unreduced invariants 
are shown in Table IV.

\section{The  SU(3) Invariants}

Finally, we
 look for the  set of invariants which are  invariant under the full $SU(3)$  that correspond to the set of gauge-fixed invariants that we have found in Tables III, IV.
 
 The gauge fixed invariants provide some clues as to the form of the invariants that we need. In particular, the flavor symmetry is very constrained by the symmetry of the gauge fixed invariants and the number of $V^0$ that they contract with, because the $V^0$ must themselves be symmetrized. Therefore, for example, the expression 
 $V_0^2\cdot V^{IJ}$ must correspond to an $SU(3)$ invariant with flavor symmetry within 
 \yng(2)$\times$ \yng(2).
 
 In  this case, though,
  we already see that we only need $SU(3)$ invariants with up to six fields. This is a small enough number that we can do a exhaustive search through all combinations of these $SU(3)$ adjoints with any flavor symmetry, and we do not need to use detailed information from the structure of the gauge fixed invariants.

To look for these invariants, we use the program LIE~\cite{LIE}.
The program can calculate the decomposition into $SU(3)$ representations of any combinations of adjoints with any flavor symmetry. We use it to compute the 
 the number of invariants of any combination  of up to seven $SU(3)$ adjoints.
 
We must however note that many  invariants found by LIE are not independent, but are rather products of smaller invariants.
 For example from LIE, we find one invariant with two fields with the symmetry \yng(2). With four fields, we find two invariants with symmetry \yng(2,2). However, we know that we can take a product of two of the two-field invariants which has the symmetry  \yng(2,2).
 Subtracting this off, we find one new independent invariant with symmetry \yng(2,2).

We repeat this process for all possible combinations of up to seven adjoints,  as shown in the tables in Appendix 1. Nonzero entries in the final column correspond to independent invariants. As expected, 
we find no independent invariants with 7 fields.
Explicit expressions for these invariants are shown in Table \ref{su3invts}.

The computation in Appendix 1 by itself cannot rule out the existence of invariants with eight or more fields. However, if we can reproduce all the gauge-fixed invariants, then we are assured that there are no further independent invariants.
We check this directly, by  gauge 
fixing one adjoint  to the form (\ref{X0form}) and attempting to reproduce all gauge-fixed invariants in Table \ref{genericgaugefixedinvts}. Similarly, we can set $b=c$, and see if all the gauge-fixed invariants in Table \ref{enhancedgaugefixedinvts} can be reproduced. This indeed can be done, as shown 
in Appendix 2. This confirms that we have found all the invariants of this theory.


\begin{table}[ht]
\centering 
\begin{tabular}{|c| c| c| c| c|}
\hline\hline 
Tableaux &  $SU(3)$ Invariants  \\ [0.5ex] 
\hline 
 \rule{0pt}{1.2em}
 $\begin{array}{c}\young(IJ)\end{array}$\ & 
 $\begin{array}{c} Tr(X^IX^J) \end{array}$
 \\ 
\hline 
 \rule{0pt}{1.2em}
 $\begin{array}{c}\young(IJK)\end{array}$\ & 
 $\begin{array}{c} Tr(X^IX^JX^K) \end{array}$
 \\ 
\hline
 \rule{0pt}{2.2em}
 $\begin{array}{c}\young(I,J,K)\end{array}$\ & 
 $\begin{array}{c} Tr(X^IX^JX^K) \end{array}$ 
  \\ 
\hline
 \rule{0pt}{1.8em}
 $\begin{array}{c}\young(IK,JL)\end{array}$\ & 
 $\begin{array}{c} Tr([X^IX^J][X^KX^L])
 \end{array}$ 
  \\ 
\hline
 \rule{0pt}{2.2em}
 $\begin{array}{c}\young(IL,J,K)\end{array}$\ & 
 $\begin{array}{c} Tr([X^I,X^J,X^K]X^L) \end{array}$ 
   \\ 
\hline
 \rule{0pt}{2.2em}
 $\begin{array}{c}\young(ILM,J,K)\end{array}$\ & 
 $\begin{array}{c} 
 Tr([X^I,X^J,X^K]\{X^LX^M\}) 
 \end{array}$ 
    \\ 
\hline
 \rule{0pt}{2.2em}
 $\begin{array}{c}\young(IL,JM,K)\end{array}$\ & 
 $\begin{array}{c} Tr([X^IX^JX^K][X^LX^M])
\end{array}$   
     \\ 
\hline
 \rule{0pt}{3.4em}
 $\begin{array}{c}\young(I,J,K,L,M)\end{array}$\ & 
 $\begin{array}{c} Tr([X^IX^JX^KX^LX^M]) \end{array}$  
      \\ 
\hline
 \rule{0pt}{1.8em}
 $\begin{array}{c}\young(IKM,JLN)\end{array}$\ & 
 $Tr([X^I,X^J][X^K,X^L][X^M,X^N])$
       \\ 
\hline
 \rule{0pt}{2.8em}
 $\begin{array}{c}\young(IMN,J,K,L)\end{array}$\ & 
 $\begin{array}{c} Tr([X^IX^JX^KX^L]\{X^MX^N\})
\end{array}$
 \\
\hline
\end{tabular}
\caption{This table shows explicit expressions for an independent set of $SU(3)$ invariants with adjoints.}
\label{su3invts}
\end{table}

\section{Conclusions}

We have discussed a method to determine a set of independent  invariants of a theory with a general gauge group and matter field in a general representation.
We have done this by using a theorem  that relates gauge-fixed configurations
to the independent invariants in a gauge theory. 
Specifically, this theorem asserts that the constant configurations of the fields, identified
by complex gauge transformations, are in one-to-one correspondence with the invariants in the theory.
We have shown that this method provides a straightforward approach
 to find the independent invariant tensors. 

We
have  applied these methods to a theory with
$SU(3)$ gauge symmetry and  with matter in the adjoint, and found the independent invariant tensors, listed in Table \ref{su3invts}.

Many other groups and representations remain to be studied; 
in particular little is known of the invariant tensors of the 
exceptional groups. 
We hope to return to this topic in future work.

\

\

\begin{center}
{\bf Acknowledgements}
\end{center}
This work was supported in part by NSF Grant No.~PHY-1915005.

\newpage

\

\begin{center}
{\bf Appendix 1}
\end{center}

This appendix summarizes the calculations to establish all independent invariant structures in $SU(3)$ with adjoints.

For each candidate tableau in the first column (shown as a list of the number of boxes per row) we use the program LIE to find the number of invariants (shown in the second column) with that symmetry .We then subtract off any 
invariants that can be written as a product of smaller invariants; these are shown in the following column of the table. The net number of independent invariants after subtraction is shown in the final column. 

On occasion the number to subtract may be larger than the number in the second column; this indicates the existence of a relation between the invariants to be subtracted. We have not explored these relations in detail.

\begin{tabular}{|c| c| c| c| c|} 
\hline\hline 
Tableau  & \# Invts & Subtract & Net  \\ [0.5ex] 
\hline 
(2)  & 1 &- &1 \\ 
\hline
(1,1) & 0 &- &0 \\ 
\hline 
\hline

(3)  & 1 &- &1 \\ 
\hline
(2,1) & 0 &- &0 \\ 
\hline
(1,1,1) & 1 &- &1 \\ 
\hline
\hline

(4) &  1 &(2)(2) &0 \\ 
\hline
(3,1)  & 0 &- &0 \\ 
\hline
(2,2)  & 2 &(2)(2) &1 \\ 
\hline
(2,1,1)  & 1 &- &1 \\ 
\hline
(1,1,1,1) &  0 &- &0 \\ 
\hline
\hline

(5)  & 1 &(2)(3) &0 \\ 
\hline
(4,1) & 1 &(2)(3) &0 \\ 
\hline 
(3,2)  & 1 &(2)(3) &0 \\ 
\hline
(3,1,1)  & 2 &(1,1,1)(2) &1 \\ 

\hline
(2,2,1) &  1 &- &1 \\ 
\hline
(2,1,1,1)  & 1 &(1,1,1)(2) &0 \\
\hline
(1,1,1,1,1)  & 1 &- &1 \\
\hline
\hline

(6) \ & 2 &
$\begin{array}{c}(2)(2)(2)
\\
(3)(3)\end{array}$
 &0 \\ 
\hline 
(5,1)   & 0 &- &0 \\ 
\hline
(4,2)  & 3 &$\begin{array}{c}(2)(2)(2)
\\
(2,2)(2)
\\
(3)(3)\end{array}$ &0 \\ 
\hline
(3,3)  & 1 &- &1 \\ 
\hline
(4,1,1)   & 2 &$\begin{array}{c}
(1,1,1)(3)\\
(2,1,1)(2)\end{array}$&0 \\
\hline
(3,2,1)  
 &  2 &$\begin{array}{c}
(2,2)(2)
\\
(2,1,1)(2)\end{array}$ &0 
\\
\hline
(2,2,2) &  3 &
$\begin{array}{c}
(2)(2)(2)
\\
(2,2)(2)
\\
(1,1,1)(1,1,1)
\end{array}$
&0
\\
\hline
(3,1,1,1)  & 3 &
$\begin{array}{c}
(2,1,1)(2)
\\
(1,1,1)(3)
\end{array}$
&1
\\
\hline
(2,2,1,1)  & 1 &(2,1,1)(2) &0
\\
\hline
(2,1,1,1,1)  & 1 &(1,1,1)(1,1,1) &0
\\
\hline
(1,1,1,1,1,1)  & 0 &- &0
\\
\hline
\end{tabular}
\quad
\begin{tabular}{|c| c| c| c| c|} 
\hline\hline 
Tableau  & \# Invts & Subtract & Net  \\ [0.5ex] 
\hline 
(7)   & 1 &(3)(2)(2) &0 \\ 
\hline 
(6,1)   & 1 &(3)(2)(2) &0 \\ 
\hline
(5,2)  & 3 &$\begin{array}{c}2\times (3)(2)(2)
\\
(2,2)(3)\end{array}$ &0 \\ 
\hline
(4,3)  & 1 &
$\begin{array}{c}(3)(2)(2)
\end{array}$
&0 \\ 
\hline
(5,1,1)   &  3 &
$\begin{array}{c}(1,1,1)(2)(2)
\\
(2,1,1)(3)
\\
(3,1,1)(2)
\end{array}$
&0 \\
\hline
(4,2,1)  
  & 5 &
 $\begin{array}{c}
(2)(2,2,1)
\\
(2)(3,1,1)
\\
(3)(2,1,1)
\\
(2,2)(3)
\\
(3)(2)(2)
\end{array}$&0 
\\
\hline
(3,2,2)   & 3 &
$\begin{array}{c}(1,1,1)(2,1,1)
\\
(2,2)(3)
\\
(3)(2)(2)
\end{array}$
&0
\\
\hline
(3,3,1)  &  3 &
$\begin{array}{c}
(2)(3,1,1)
\\
(2,2)(1,1,1)
\\
(1,1,1)(2)(2)
\end{array}$
&0
\\
\hline
(4,1,1,1)   & 2 &
$\begin{array}{c}
(2)(3,1,1)
\\
(3)(2,1,1)
\\
(1,1,1)(2)(2)
\end{array}$
&0
\\
\hline
(3,2,1,1)  & 5 &
$\begin{array}{c}
(2)(2,2,1)
\\
(2)(3,1,1)
\\
(1,1,1)(2,1,1)
\\
(3)(2,1,1)
\\
(2,2)(1,1,1)
\\
(1,1,1)(2)(2)
\end{array}$
&0
\\
\hline
(2,2,2,1)&  1 &(2,1,1)(1,1,1)  &0
\\
\hline
(3,1,1,1,1)&  2 &$\begin{array}{c}(2,1,1)(1,1,1) \\
(2)(1,1,1,1,1)
\end{array}$ &0
\\
\hline
(2,2,1,1,1)
 & 2 &
$\begin{array}{c}(2,1,1)(1,1,1) \\
(2,2)(1,1,1) 
\end{array}$
&0
\\
\hline
(2,1,1,1,1,1) \ & 1 &(2,1,1)(1,1,1) &0
\\
\hline
(1,1,1,1,1,1,1)  & 0 &- &0
\\
\hline
\end{tabular}

\newpage
\begin{center}
{\bf Appendix 2}
\end{center}

In this Appendix, we show that the $SU(3)$ invariants in Table V are sufficient to reproduce all the invariants in Tables III, IV. We replace various field in the $SU(3)$ invariants with the gauge fixed value (\ref{X0form}). The resulting expression is shown in the corresponding second column. It can be seen that all the gauge fixed invariants can be reproduced from this replacement.We have done this separately for the generic points of parameter space in the following table  and for the point of enhanced symmetry in the table on the following page.
\begin{table}[ht]
\centering 
\begin{tabular}{|c| c| c| c| c|} 
\hline
$SU(3)$ invariant & Expression after gauge fixing 
&
$SU(3)$ invariant &  Expression after gauge fixing\\ [0.5ex] 
\hline 
\rule{0pt}{1.2em}
$\begin{array}{c}\young(0I)
\\
\young(00I)
\end{array}$ 
&
$\begin{array}{c}V_0\cdot V^{I}
\\
V_0^2\cdot V^{I}
\end{array}$

&
\rule{0pt}{2.4em}
$\begin{array}{c}\young(IK,J,0)\end{array}$
-$\begin{array}{c}\young(IJ,K,0)\end{array}$

&$
-V_0\diamond (V^{'IJK}+W^{'IJK})
+...$

\\
\hline
\rule{0pt}{1.2em}$\begin{array}{c}\young(IJ)
\\
\young(0IJ)
\\
\young(00,IJ)
\end{array}$
&
$\begin{array}{c}Tr(V^{IJ})
\\
-2V_0\cdot V^{IJ}+...
\\
3V_0^2\cdot V^{IJ}+...
\end{array}$

&
\rule{0pt}{1.8em}
$\begin{array}{c}\young(IK,J0)\end{array}$
-$\begin{array}{c}\young(IJ,K0)\end{array}$
&
$
 V_0 \diamond ((3/4) V^{'IJK}
 -(1/4) W^{'IJK})
 +...$

\\
\hline
\rule{0pt}{2.4em}
$\begin{array}{c}
\young(0,I,J)
\\
\young(00,I,J)
\\
\young(000,I,J)
\end{array}$ 
&
$\begin{array}{c}6V_0 \diamond V^{'IJ}+...
\\
2V_0^2 \diamond V^{'IJ}+...
\\
-2V_0^3 \diamond Tr(V^{'IJ})+..
\end{array}$

&
\rule{0pt}{2.4em}
$\begin{array}{c}\young(IK,J0,0)\end{array}$
-$\begin{array}{c}\young(IJ,K0,0)\end{array}$
&
$
(3/2)V_0^2\diamond (V^{'IJK})
-(1/2)V_0^2\diamond (W^{'IJK})+...$

\\
\hline
\rule{0pt}{1.2em}
$\begin{array}{c}\young(IJK)
\end{array}$ 
&
$Tr(V^{IJK})+...$

&
\rule{0pt}{2.4em}
$\begin{array}{c}\young(IK0,J,0)\end{array}$
-$\begin{array}{c}\young(IJ0,K,0)\end{array}$
&
$
(1/2)V_0^2\diamond (V^{'IJK}+W^{'IJK})+...$

\\
\hline
\rule{0pt}{2.4em}
$\begin{array}{c}\young(I,J,K)\end{array}$ 
&
$Tr(W^{IJK})+...$

&
\rule{0pt}{1.8em}
 $\begin{array}{c}
 \\
\young(IJK,000)
\end{array}$
&
$V_0^3 \diamond Tr(V^{'IJK})+...$

\\
\hline
\rule{0pt}{1.8em}
$\begin{array}{c}\young(IK,J0)\end{array}$
+$\begin{array}{c}\young(IJ,K0)\end{array}$
&
$3 V_0 . (V^{IJK}+W^{IJK})
+...
$

&

\rule{0pt}{3.2em}
$\begin{array}{c}\young(000,I,J,K)\end{array}$ 
&
$V_0^3 \diamond Tr(W^{'IJK})+...$

\\

\hline
\rule{0pt}{2.4em}
$\begin{array}{c}\young(IK,J,0)\end{array}$
+$\begin{array}{c}\young(IJ,K,0)\end{array}$

&$(3/2)V_0.V^{IJK}
-(1/2)V_0.W^{IJK}
+...
$
 & &
\\
\hline
\rule{0pt}{2.4em}
$\begin{array}{c}\young(IK0,J,0)\end{array}$
+$\begin{array}{c}\young(IJ0,K,0)\end{array}$
&
$
-(3/2)V_0^2V^{IJK}-(1/2)V_0^2W^{IJK}
+...
$
 & &
\\
\hline
\rule{0pt}{2.4em}
$\begin{array}{c}\young(IK,J0,0)\end{array}$
+$\begin{array}{c}\young(IJ,K0,0)\end{array}$
&
$3V_0^2.(V^{IJK}+W^{IJK})
+...
$
 & &
\\
\hline
\end{tabular}
\caption{This shows the matching of the $SU(3)$ invariants to the gauge-fixed invariants at the generic point of the theory.}
\end{table}

\begin{table}[h]
\centering 
\begin{tabular}{|c| c| c| c| c|} 
\hline\hline 
$SU(3)$ invariant & Expression after gauge fixing \\ [0.5ex] 
\hline 
\rule{0pt}{1.2em}

$\begin{array}{c}\young(0I) \end{array}$ 
&$4s^{I}$
\\
\hline
 \rule{0pt}{1.8em}
$\begin{array}{c}\young(IJ) \end{array}$  
&
$2a^K_i a^L_i+2q^{I}.(Q)^{J}+2Q^{I}.q^{J}+...$
\\
\hline
 \rule{0pt}{1.8em}
 $\begin{array}{c} \young(0IJ)\end{array}$ 
&
$2a^I_i a^J_i-q^{I}.(Q)^{J}-Q^{I}.q^{J}+...$
\\
\hline
\rule{0pt}{1.8em} $\begin{array}{c} \young(0,I,J)\end{array}$ 
&
$-3q^{I}.Q^{J}+3Q^{I}.q^{J}+...$

\\
\hline
\rule{0pt}{2.0em}
$\begin{array}{c}\young(IJK)\end{array}$ 
&
$q^{I}\sigma^i Q^{J}a^K_i +\{IJK\}+...$
\\
\hline
\rule{0pt}{2.0em}  
$\begin{array}{c}\young(IK,J0)\end{array}$  
&
$-q^{I}\sigma^i Q^{K}a^J_i
+
q^{K}\sigma^i Q^{I}a^J_i
+q^{J}\sigma^i Q^{K}a^I_i
-
q^{K}\sigma^i Q^{J}a^I_i+...$
\\
\hline
\rule{0pt}{2.0em}
$\begin{array}{c}\young(I,J,K)\end{array}$ 
&
$
q^{J}\sigma^i Q^{K}a^I_i
+{1\over 2}a^Ia^Ja^K+[IJK]+...$
\\
\hline
\rule{0pt}{2.0em}
$\begin{array}{c}\young(I0,J,K)\end{array}$ 
&
$-(4/3)
q^{J}\sigma^i Q^{K}a^I_i
+(1/3)\epsilon^{ijk}a_i^Ia_j^Ja_k^K +[IJK]+...$

\\
\hline
\rule{0pt}{3.0em} $\begin{array}{c}\young(IK,JL)\end{array}$ 

&
$-q^Ia^Ka^LQ^J-q^Ka^Ia^JQ^L
-q^Ia^Ja^KQ^L-q^Ka^La^IQ^J+([IJ][ KL])+...$

\\
\hline
\rule{0pt}{3.0em}
$\begin{array}{c}\young(IL,J,K)\end{array}$ 
&
$
-q^Ia^Ja^KQ^L
+
q^La^Ia^JQ^K
+q^Ka^La^IQ^J
 +q^Ja^Ka^L Q^I+[IJK]$
$ 
q^{I}\sigma^{ij}  Q^{J} a^K_i a^L_j 
$

\\
\hline

$\begin{array}{c}\young(0,I,J,K,L)
\end{array}$

&
\rule{0pt}{3.0em}
$ 2q^Ia^Ja^KQ^L+[IJKL]$
$\begin{array}{c}
q^{I}\sigma^{ij}  Q^{J} a^K_i a^L_j 
\end{array}$ 

\\
\hline
\rule{0pt}{3.0em}
$\begin{array}{c}\young(IKL,J,0)\end{array}$
&$q^L(a^Ia^J)Q^K+q^K(a^Ia^J)Q^L$,
$q^{I}\sigma^{ij}  Q^{J} a^K_i a^L_j $ 
\\
\hline
\end{tabular}
\caption{This shows the matching of the $SU(3)$ invariants to the gauge-fixed invariants at the point of enhanced symmetry.}
\end{table}

\newpage
\

\end{document}